\documentclass[12pt]{iopart}
\begin{document}

\title[Pollaczek polynomials and hydrogen radial Schr\"odinger
  equation]{Explicit representations of Pollaczek polynomials
  corresponding to an exactly solvable discretisation of hydrogen
  radial Schr\"odinger equation}

\author{Matias Aunola}

\address{Defence Forces Technical Research Centre (PvTT), P.O. Box
10, FIN-11311 Riihim\"aki, Finland}

\begin{abstract}
  We consider an exactly solvable discretisation of the radial
  Schr\"odinger equation of the hydrogen atom with $l=0$. We first
  examine direct solutions of the finite difference equation and
  remark that the solutions can be analytically continued entire
  functions. A recursive expression for the
  coefficients in the solution is obtained. The next step is to
  identify the related three-term recursion relation for Pollaczek
  polynomials. One-to-one correspondence between the spectral and
  position representations facilitates the evaluation of Pollaczek
  polynomials corresponding to the discrete spectrum. Finally, we
  obtain two alternative and explicit expressions for the solutions of
  the original difference equation. 
 
\end{abstract}


\submitto{\JPA}


\section{Introduction\label{sec:intro}}

The general solution of the Schr\"odinger equation, the corresponding
energy levels and resulting atomic shell model are taught to all
students of quantum mechanics. Experimental and theoretical studies of
Rydberg atoms with a single electron have illuminated the wave
functions and the classical limit of quantum mechanics.

Here, we will concentrate on a specific aspect of theoretical studies,
namely discretisations and orthogonal polynomials.  Orthogonal
polynomials, namely Laguerre polynomials, are already present in the
solution of the Schr\"odinger equation.  The subject of orthogonal
polynomials related to the hydrogen atom was reviewed in 1991 by
Dehesa et al.~in Ref.~\cite{deh91}.  Apart from polynomials present in
the solutions, different kinds of discretisations induce further
orthogonal polynomials. The most obvious ones arise from direct
discretisations of the Schr\"odinger equation.  Unfortunately, the
explicit form of these polynomials is not usually known. The continuum
states and $L^2$ discretisations of the continuum have been studied
already in the 1970's \cite{yam75,bro78}.  Recently, a corresponding
solution for the Dirac-Coulomb problem was presented in
Ref.~\cite{alh04}. In the context of condensed matter physics
discretised Schr\"odinger equations can also be used as simple models
of nanoscale systems \cite{boy04} or related to the tight-binding
approximation \cite{hay72}.

In this paper, we show that the $l=0$ states of the hydrogen atom can
be exactly and explicitly obtained for the symmetric discretisation of
the second-order derivative. First this is done by inserting an
explicit ansatz in to the difference equation. Later on, we relate the
difference equation to the three-term recursion relation for Pollaczek
polynomials with specific parameters. Finally, a surprisingly simple
and explicit expression is obtained for Pollaczek polynomials and the
corresponding solutions of the discretised Schr\"odinger equation.

\section{Initial steps}

Let us consider the radial Schr\"odinger equation for the hydrogen
atom for the hydrogen atom, i.e.
\begin{equation}
-\frac{\hbar^2R''(r) }{2m}-\frac{\hbar^2 R'(r)}{m r}-\frac{e^2 R(r)}
{4\pi\varepsilon_0 r}+\frac{\hbar^2l(l+1)R(r)}{2mr^2}=E R(r). 
\end{equation}
In so-called natural units and for $l=0$ the equation simplifies to
\begin{equation}
-\frac{u''(r)}{2}-\frac{u(r)}r=E u(r),
\end{equation}
where $u(r)=r R(r)$. The simple eigenvalues and well-known solutions
are expressible in terms of associated Laguerre polynomials of the
first kind
\begin{equation}
\fl u_n(r)=r L_{n}^{1}(2r/n) e^{-r/n}=e^{-r/n}\sum_{k=1}^n \frac{(-2/n)^{k-1}}
{k!}\left(\matrix{ n-1\cr k-1}\right)r^k,\quad
E_n=-\frac1{2n^2},\label{eq:solu0}
\end{equation}
where $n=1,2,\ldots$. 
Next we discretise this equation using the symmetric second-order difference
and obtain a finite-difference equation 
\begin{equation}
-\frac{u(r-\delta)-2u(r)+u(r+\delta)}{2\delta^2}-\frac{u(r)}{r}=E(\delta)
u(r).\label{eq:difference0}
\end{equation}
In the following section, we lift the restrictions that $r$ and
$\delta$ must lie on the positive axis and allow complex values for
both. Of course, any solution of Eq.~(\ref{eq:difference0}) multiplied
by an arbitrary function with period $\delta$ is still a solution.
Nevertheless, we will concentrate on solutions and eigenvalues that
tend to corresponding classical solutions in the limit $\delta
\rightarrow 0$.

\section{Solutions in coordinate representation}

For nonzero values of $\delta$ we note that functions $u(r)$ 
are also solutions of
\begin{equation}
{u(r-\delta)/2+u(r+\delta)}/2+\delta^2 u(r)/r=\mu
u(r),\label{eq:difference}
\end{equation}
where $\mu=-\delta^2E+1$. This problem has been studied by Berezin
in the case of purely imaginary $\delta$ in Ref.~\cite{ber97}.
We discovered the existence of an explicit solution in Ref.~\cite{aun03}.
Here a more transparent and instructive derivation is given and
we are able to obtain a new, recursive formula for arbitrary
terms in the solution. 

Let us insert an ansatz 
\begin{equation}
  u(r)=e^{\beta r}\sum_{k=1}^n\alpha_k r^k,\label{eq:ansatz}
\end{equation}
into Eq.~(\ref{eq:difference}) and assume $\alpha_k\ne0$, which
obviously corresponds to the solution $u_n(r)$ in
Eq.~(\ref{eq:solu0}). Now the equations must hold identically in $r$
so each coefficient of $r^j$ must vanish. This yields
\begin{equation}
\fl  \sum_{k=\max(1,j-1)}^n\gamma_{j,k}\alpha_k=0,
\ \ \gamma_{j+1,k}:=\delta^{k-j}\left(\matrix{
    k\cr j}\right)(e^{\beta\delta}+
   (-1)^{k-j}e^{-\beta\delta})/2
      -\mu\gamma_{k,j}+\delta^2\gamma_{k,j+1}
\end{equation}
for $ j=1,\ldots,n+1$.  The $(n+1)$th equation requires that
$\alpha_n[\cosh(\delta\beta) -\mu]=0$, so we find
$\mu=\cosh(\delta\beta)$.  Next equation is then
simplified to $\alpha_n[n\delta^{-1}\sinh(\delta\beta)+1]=0$.
These equations now yield
\begin{equation}
\mu_n(\delta)=\sqrt{1+(\delta/n)^2},\quad
\beta_n(\delta)=-\mathrm{arsinh}(\delta/n)/\delta
\label{eq:simpresult}
\end{equation}
in agreement with Ref.~\cite{aun03}. The classical limits
$E_n(\delta)\rightarrow -1/2n^2$ and $\beta_n(\delta)\rightarrow -1/n$
are also satisfied. The remaining $n-1$ equations can be used in order
to solve the constants $\alpha_k$ with $k=1,\ldots, n-1$. Note that
the present derivation is both simpler and more exhausting than the
previous one, which was based on an intelligent guess concerning the
identity of terms in the power series used. All eigenvalues and
constants in the exponential part are identical to those given in
Eq.~(\ref{eq:simpresult}).

The general solution to Eq.~(\ref{eq:difference}) now becomes 
\begin{equation}
\fl u_n^{(\delta)}(r)=\left(\sum_{k=1}^{n}\ell_k^{(n)}\,\alpha^{(n,\delta)}_k
\, r^k\right)\exp(-r\,\mathrm{arsinh}(\delta/n)/\delta),\quad
\ell_k^{(n)}:=\frac{(-2/n)^{k-1}}{k!}\left(
\matrix{n-1\cr k-1}\right).
\end{equation}
The coefficients $\{\alpha^{(n,\delta)}_k\}$ are of the form
\begin{equation}
\alpha^{(n,\delta)}_{n-k}=(1+\delta^2/n^2)^{(k-2k')/2}\sum_{m=0}^{k'}
\alpha^{(n)}_{n-k,m}\delta^{2m},
\end{equation}
where $\alpha^{(n)}_{k,0}=1$ and $k'=\lfloor k/2\rfloor$, i.e.
$k/2$ if $k$ is even and $(k-1)/2$ if $k$ is odd.
In addition, we find
\begin{equation}
\alpha^{(n)}_{n-k,m}=n^{-2m}\left(
\matrix{\lfloor k/2\rfloor \cr m}\right)
\frac{P(2m-1)(n-k)!}{(n-k+2m-1)!},
\end{equation}
where $P(2m-1)$ is a polynomial of order $2m-1$ such that 
the coefficient of $x^{2m-1}$ is equal to unity.
The general form of the leading terms is given by
\begin{eqnarray}
\alpha^{(n,\delta)}_{n}&=&1\\
\alpha^{(n,\delta)}_{n-1}&=&\sqrt{1+\delta^2/n^2}\\
\alpha^{(n,\delta)}_{n-2}&=&1+\frac{(3n-1)\delta^2}{3n^2(n-1)}\\
\alpha^{(n,\delta)}_{n-3}&=&\sqrt{1+\delta^2/n^2}
\left(1+\frac{n\delta^2}{n^2(n-2)}\right)\\
\alpha^{(n,\delta)}_{n-4}&=&1+\frac{2(n-1)\delta^2}{n^2(n-3)}+
\frac{(15n^3-30n^2+5n+2)\delta^4}{15n^4(n-1)(n-2)(n-3)}
\end{eqnarray}
Here $n$ is an arbitrary
state index, so once $\alpha^{(n,\delta)}_{n-k}$ has been
obtained, we have exact expressions for $k+1$ leading 
polynomial terms in any eigenfunction.
We have now obtained explicit expressions for
$\alpha^{(n,\delta)}_{n-k}$ when $k\le 125$ and $\alpha^{(n)}_{k,m}$
for $m\le23$ with arbitrary $k$. In order to do this the underlying
symmetries of thecoefficients have to be exploited efficiently.
Some of this work was already done in  Ref.~\cite{aun03}, and
one can find explicit values of terms up to $k\le49$ and $m\le10$
in the addendum.

We proceed by noting that the innermost
coefficients, $\{\alpha^{(n)}_{n-k,m}\}$, 
yield a general solution to the present difference equation.
We define
\begin{equation}
C_{n,k,l}:=
\frac{(-n/2)^k n!}{k!l!(n-k-l)!}\prod_{m=1}^k(n-m),\qquad
\end{equation}
and assume that all coefficients up to level
$\alpha^{(n)}_{n-(k-1),m}$ are known. The requirement that the
coefficient for each power of $\delta$ cancels separately and some
algebra yields explicit expressions for the coefficients on the next
level.  For even values of $k$ we find
\begin{eqnarray}
\fl \alpha^{(n)}_{n-k,m\ge1}=\left(-\sum_{l=k'-m}^{k'-1}C_{n,2l,k+1-2l}
\alpha^{(n)}_{n-2l,l+m-k'}/n+\sum_{l=k'-m-1}^{k'-1}C_{n,2l+1,k-2l}
\alpha^{(n)}_{n-2l-1,l+m-k'+1}\right.\cr
+\ \ \left.\left.\sum_{l=k'-m}^{k'-1}C_{n,2l+1,k-2l}
\alpha^{(n)}_{n-2l-1,l+m-k'}/n^2\right)\right/(C_{n,k,1}/n-C_{n,k,0}).
\label{eq:receven}
\end{eqnarray}
Correpondingly for odd values of $k$ this yields
\begin{eqnarray}
\fl \alpha^{(n)}_{n-k,m\ge1}&=&\left(-\sum_{l=k'-m}^{k'-1}C_{n,2l+1,k-2l}
\alpha^{(n)}_{n-2l-1,l+m-k'}/n\right.\cr
&&+\ \ \left.\left.\sum_{l=k'-m}^{k'}C_{n,2l,k+1-2l}
\alpha^{(n)}_{n-2l,l+m-k'}\right)\right/(C_{n,k,1}/n-C_{n,k,0}).
\label{eq:recodd}
\end{eqnarray}
Here $\alpha^{(n)}_{k,0}=1$ and impossible coefficients are taken to
be zero. The recursive equations do yield the general solution, but
the complexity of equations grows at an exponential rate. Thus, it
has been necessary to use even higher order symmetries to
reduce the number of equations and reach the present order 
$k=125$. All calculations have been performed using the
symbolic mathematical software \textsc{Mathematica}.

By restricting the allowed values of $r$ to the set
$\{k\delta\}_{k=1}^{\infty}$ we transform the problem into an
eigenvalue problem for an infinite tridiagonal matrix. By denoting
$u^n_k:=u_n^{(\delta)}(k\delta)$, we see that the vector
$\{u_k^n\}_{k=1}^\infty$ is an eigenvector of the matrix 
\begin{equation}
H_{kk}=\delta/k,\quad H_{k,k+1}=H_{k+1,k}=1/2,\qquad k=1,2,\ldots
\label{eq:matrix}
\end{equation}
corresponding to the eigenvalue $\mu_n=\sqrt{1+(\delta/n)^2}$.
In addition, the exponential part of the solution simplifies to 
\begin{equation}
\exp(-k\delta\,\mathrm{arsinh}(\delta/n)/\delta)=(\sqrt{1+(\delta/n)^2}
-\delta/n)^k.\label{eq:expot}
\end{equation}
Results from numerical diagonalisation agree with our results within
numerical precision, as long as convergence can be reached.  For
$\delta$ real and positive, normalised eigenvectors form an
orthonormal basis of $\ell^2$.

The polynomial part of the eigenvectors define a discretised version
of the corresponding associated Laguerre polynomials $L_n^1$. It is
not clear whether a three-term recursion relation exists for these
polynomials.

\section{Solutions in spectral representation}

The polynomial character of the exponential part shown in 
Eq.~(\ref{eq:expot}) also contains both the eigenvalue $\mu$ and
the discretisation parameter, which indicates that we should 
also examine the problem with respect to the spectral variable.
It turns out that the problem at hand corresponds to a special
case of the Pollaczek polynomials \cite{chi78}. They satisfy the three-term
recursion relation 
\begin{equation}
\fl (j+1)P_{j+1}^\lambda(x;a,b)=2[(j+\lambda+a)x+b]P_{j}^\lambda(x;a,b)
-(j+2\lambda-1)P_{j-1}^\lambda(x;a,b),
\end{equation}
where $j>0$, and initial conditions
\begin{equation}
P_{0}^\lambda(x;a,b)=1,\quad P_{1}^\lambda(x;a,b)=2(\lambda+a)x+b.
\end{equation}
From Eq.~(\ref{eq:matrix}) we obtain the recursion relation
\begin{equation}
(j+1)u_{j+1}=2[(j+1)x-\delta]u_j-(j+1)u_{j-1},\quad j(=k-1)>0,
\end{equation}
and identify the parameters $\lambda=1$, $a=0$ and $b=-\delta$.
The discrete spectrum agrees with the calculations performed above, i.e.
\begin{equation}
x_m=\sqrt{1+\delta^2/(m+1)^2}, \quad m=0,1,\ldots.
\label{eq:spectrumdis}
\end{equation}
We have not yet studied the absolutely continuous spectrum in the
range $[-1,1]$.

The explicit formula for Pollaczek polynomials reads
\begin{equation}
P_n^\lambda(\cos\theta;a,b)=\sum_{k=0}^n\frac{(-\lambda+i\Phi(\theta))_k
(\lambda+i\Phi(\theta))_{n-k}}{k!(n-k)!}e^{i\theta(2k-n)},
\label{eq:polexplicit}
\end{equation}
where $x:=\cos\theta$, $\Phi(\theta)=(a\cos\theta+b)/\sin\theta$ and
$(A)_k=A(A+1)\cdots(A+k-1)$. The orthogonality of Pollaczek polynomial
is defined with respect to the interval $[-1,1]$, where $\cos\theta$
and $\sin\theta$ are easily defined. Here, it would be tempting to use
\begin{equation}
\sin\theta=\pm i\delta/(m+1),
\end{equation}
which yields simple terms to be inserted in
Eq.~(\ref{eq:polexplicit}).  Nevertheless, the correct way to do this
as well as the corresponding interpretation are not obvious to us.

 Below, we obtain a simpler way to
express the polynomials $P_j(x):=P_{j}^1(x;0,-\delta)$ for
$x(>1)$ within the discrete spectrum~(\ref{eq:spectrumdis}).
The results of the previous section show that we can write
\begin{equation}
P_j(x_m)=(x_m-\delta/(m+1))^{j-m}Q_j^m(x_m),
\end{equation}
where $Q_j^m$ is a polynomial of degree $m$. By extracting the
polynomials $Q_j$, we can reconstruct the corresponding Pollaczek
polynomials. The next step is to evaluate polynomial relations
with respect to the index $j$, i.e. express the coefficients
of the polynomials in as functions of $j$. Thus, we write
\begin{eqnarray}
P_j(x_m)&=&(j+1)\sum_{l=0}^j\left(\matrix{ j\cr l}\right)
(x_m)^{j-l}\left(-\frac{\delta}{m+1}\right)^l\beta_{m,l}\\
Q_j(x_m)&=&(j+1)\sum_{l=0}^m\left(\matrix{ m\cr l}\right)
(x_m)^{m-l}\left(-\frac{\delta}{m+1}\right)^l\gamma_{j,l}.
\end{eqnarray}
Both factors $\beta_{j,l}$ and $\gamma_{m,l}$ appear to be quite
complicated at first. Some general features can be gleaned out, but
the breakthrough is achieved in three steps. First is the observation
that $\gamma_{j,l}=\beta_{j,l}$ and the next amounts to the symmetry
$\beta_{j,l}=\beta_{l,j}$. Finally, we find
\begin{equation}
\beta_{j,m}=\sum_{l=0}^{\min(j,m)}\frac{2^l}{l+1}
\left(\matrix{ j\cr l}\right)
\left(\matrix{ m\cr l}\right).
\end{equation}
This is the solution we have been looking for, explicitly:
\begin{eqnarray}
\fl P_j(x_m)=(j+1)\sum_{l=0}^j
(x_m)^{j-l}\left(\frac{-\delta}{m+1}\right)^l
\left[\left(\matrix{ j\cr l}\right)
\sum_{k=0}^{\min(m,l)}\frac{2^k}{k+1}
\left(\matrix{ m\cr k}\right)
\left(\matrix{ l\cr k}\right)\right]\cr
\fl=(j+1)\left(x_m-\frac{\delta}{m+1}\right)^{j-m}\sum_{l=0}^m
(x_m)^{m-l}\left(\frac{-\delta}{m+1}\right)^l
\left[\left(\matrix{ m\cr l}\right)
\sum_{k=0}^{\min(j,l)}\frac{2^k}{k+1}
\left(\matrix{ j\cr k}\right)
\left(\matrix{ l\cr k}\right)\right]\label{eq:solution}
\end{eqnarray}
The first expression is more convenient for $j\le m$, while the second
is more compact for $j>m$. Note that expressions are manifestly
identical for $j=m$.  The unnormalised, general solution to the matrix
eigenvalue problem~(\ref{eq:matrix}) now simplifies to
\begin{equation}
u_k^n=P_{k-1}(x_{n-1}), \quad \mu_n=x_{n-1}.\label{eq:pollsolu}
\end{equation}
For real values of $\delta$ the solutions satisfy the orthogonality
relation
\begin{equation}
\sum_{k=1}^\infty u_k^n u_k^{n'}\propto \delta_{n,n'}
\end{equation}
and normalisation requires that
\begin{equation}
\sum_{k=1}^\infty \vert u_k^n\vert^2=1.
\end{equation}

For small values of $j$ or $m$ we can be sure that Pollaczek
polynomial do satisfy the recursion relation.  The solutions for
spectral and coordinate representations are identical because they are
solutions to the same difference equation. Thus, in the limit
$\delta\rightarrow 0$, the Pollaczek polynomials tend to Laguerre
polynomials with corresponding exponential parts, even if the
variables $x$ and $r$ do not coincide. The extraction process
guarantees that each term has been uniquely and correctly identified
with corresponding powers of $x_m$ and $-\delta/(m+1)$.  Next, we
identified several sets of equations that these coefficients satisfy.
This allowed us to construct further coefficients in the series
without redoing the extraction process. As the final step, the
explicit expressions were conjectured and verified against known
results.  The general solution to Eq.~(\ref{eq:difference}) can be
considered as a partially proven conjecture.\footnote{The required
  intermediate steps are available at request from
  Matias.Aunola@pvtt.mil.fi}

\section{Discussion\label{sec:conclu}}

We have examined a simple discretisation of the radial Schr\"odinger
equation and shown that it is exactly solvable. We were able to obtain
an explicit solution both in terms of the radial coordinate as well as
the spectral variable, i.e. the eigenvalue. In future, one can simply
use the existing solution, e.g.~in the form of a piece of computer
code based on Eq.~(\ref{eq:solution}). Simultaneously, we derived
simple expressions to Pollaczek polynomials $P_{j}^1(x;0,-\delta)$ for
the discrete mass points $x_m$. Initial steps of the present approach
are due to earlier research on the discretised 1D harmonic oscillator,
where we obtained asymptotical representations of Mathieu functions
\cite{aun03b}. Much work still remains and alternative approaches
should be applied to these problems.

As a final note, we state that the present discretisation of the
Schr\"odinger equation can also be used when visualising hydrogen
radial wave functions.  The possibility of comparing a numerical
algorithm against exact results is not too common, especially if exact
results are for the algorithm itself.  Of course, it is not possible
to carry the recursion either to infinite order or with infinite
precision.

\ack

Illuminating discussions with Dr. T. Hyt\"onen, Prof. M. Ismail and
Prof. R. Askey are gratefully acknowledged. The author thanks
Dr.~J.~Merikoski for his comments on the final draft.

\end{document}